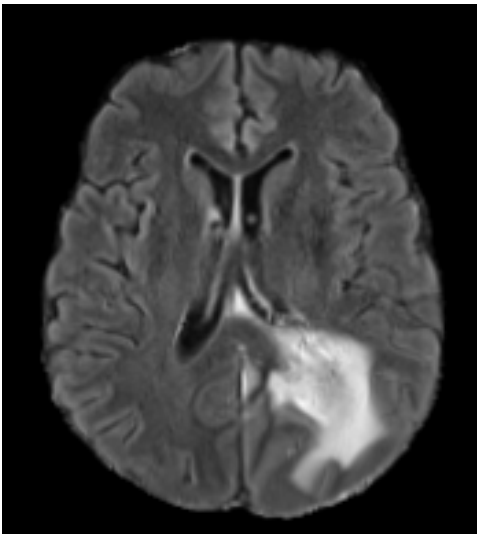

# AI Solution for Effective Diagnosis, Prognosis and Treatment Planning for Brain Tumor Patients

Vikram Goddla

Detroit Country Day School

# Table of Contents

## Abstract

Glioblastomas are the most common malignant brain tumors in adults[1,2]. Approximately 200,000 people die each year from Glioblastoma in the world[30]. Glioblastoma patients have a median survival of 12 months with optimal therapy and about 4 months without treatment[1,2,27]. Glioblastomas appear as heterogenous necrotic masses with irregular peripheral enhancement, surrounded by vasogenic edema[27]. The current standard of care includes surgical resection, radiotherapy and chemotherapy, which require accurate segmentation of brain tumor subregions[1,2]. For effective treatment planning, it is vital to identify the methylation status of the promoter of Methylguanine-Methyltransferase (MGMT-Promoter), a positive prognostic factor of patient's chemotherapy response[1,2].

However, current methods for brain tumor segmentation are tedious, subjective and not scalable[1], and current techniques to determine the methylation status of MGMT-promoter involve surgically invasive procedures, which are expensive and time consuming[1]. Hence there is a pressing need to develop automated tools to segment brain tumors and non-invasive methods to predict methylation status of MGMT-promoter, to facilitate better treatment planning and improve survival rate[1,6,8,17].

I created an integrated diagnostics solution powered by Artificial Intelligence to automatically segment brain tumor subregions and predict MGMT-promoter methylation status, using brain MRI scans. My AI solution is proven on large datasets with performance exceeding current standards and field tested by board certified neuroradiologist(s), using different images from local institutions. With my solution, physicians can submit brain MRI images, and get segmentation and methylation predictions in minutes, and guide brain tumor patients with effective treatment planning and ultimately improve survival time.

## Background

A primary brain tumor is one that originates in the brain as opposed to one that starts in another part of the body and spreads (metastasizes) to the brain. There are several types of primary brain tumors, some of which cannot be assigned an exact type as their location may not be accessible for full testing. The most common brain tumors are divided into Glioma and Non-Glioma tumor types.

The exact origin of Gliomas is still unknown. However, they are thought to grow from "glial" cells or glial precursor cells. Glial cells, as defined by the word "glia" meaning glue, surround and hold neurons in place and provide supporting functions. Glioma is graded on a scale of one to four. Grade four Glioma is the most aggressive type and is also known as Glioblastoma.

Gliomas are also classified based on the type of glial cells they originate in, which include ependymal cells, astrocytes and oligodendrocytes. A tumor that starts in astrocytes is called astrocytoma. A diffuse astrocytoma is a type of glioma that has ill-defined boundaries which tend to infiltrate neighboring healthy tissue.

Glioblastoma and diffuse astrocytic glioma with molecular features of glioblastoma are the most aggressive malignant primary tumors in the central nervous system. They are typically seen well on brain MRI (Magnetic Resonance Imaging) scans. Glioblastomas usually contain various heterogenous sub regions including tumor core, enhancing tumor and edema with varying histologic and genomic phenotypes. This extreme intrinsic heterogeneity of gliomas is also portrayed in their radiographic phenotypes, as their sub regions are depicted by different intensity profiles spread across multimodal and multiparametric MRI scans, reflecting differences in tumor biology.

For several brain tumor related clinical applications, including surgical treatment planning, image-guided interventions, monitoring tumor growth and generation of radio-therapy maps, it is extremely important to accurately identify brain tumor sub regions and its boundaries. However, manual detection and tracing of brain tumor sub regions is tedious, time consuming and subjective. The manual process is carried out by the radiologists, in a clinical setup, who often rely on qualitative visual analysis of the MRI scans and hence may not be consistent across the board. Moreover, such manual analysis may also be impractical when dealing with a multitude of patients. My research developed deep learning brain tumor segmentation model that automatically produce detailed segmentation of brain tumor sub regions, that correspond to those created by neuroradiologists.

Recent studies of central nervous system tumors highlighted the appreciation of integrated diagnostics, as opposed to relying on purely morphologic and histopathologic classification[2]. Such integrated diagnostics transitioned the clinical tumor diagnosis to include molecular cytogenetic characteristics, leading to improved predictive, prognostic and diagnostic imaging biomarkers, hence yielding the benefit towards non-invasive precision medicine[2]. MGMT, methyl guanine-DNA methyl transferase, is a DNA repair enzyme. Methylation of MGMT promoter is identified as a favorable prognostic factor and a predictor of chemotherapy response.

The current standard of care treatment of Glioblastomas includes surgical resection of maximum amount of tumor, based on the geographical location of the tumor, with possible addition of localized chemotherapy, followed by the implementation of Stupp protocol, which consists of radiotherapy and adjuvant chemotherapy. Where surgical resection is not feasible, patients are often subjected to local chemotherapy treatment followed by Stupp protocol. The efficacy of the treatment varies according to the biological status of the patient's MGMT genes, along with their age and general condition.

Alkylating chemotherapy usually involves administering temozolomide, which prevents cancer cells from making new DNA and hence preventing its growth. However, patients with expression of MGMT, which is a DNA suicide repair gene, tend to resist chemotherapy, as MGMT counters temozolomide effect by repairing the DNA and hence limiting its effect. On the other hand, methylation of MGMT in its promoter region is associated with loss of MGMT expression and hence diminished DNA repair activity and consequently increasing the effectiveness of alkylating chemotherapy. Hence it is important to understand the status of MGMT to guide the patient with effective treatment planning.

My research is focused on solving the issues associated with brain tumor segmentation and detection of MGMT promoter using quick, accurate and non-invasive means. I developed an integrated diagnostic tool which provides several deep learning models (both segmentation and radiogenomic models) to accurately segment the brain tumor sub regions to aid several clinical applications, as well as predict the methylation status of MGMT promoter to predict patients' response to alkylating chemotherapy. In addition, my solution lays the ground work to study other biomarkers including IDH mutations and the presence of alpha ketogluterate and 2-hydroxygluterate.

# Magnetic Resonance Imaging (MRI) of Brain

Unlike X-ray and CT (Computed Tomography) scans, which are maps of tissue density, MRI images represent a map of proton energy within tissues. MRI scanners detect radio frequency signals emitted by protons and then transforms them into an image. MRI is a very effective means to obtain exquisite images of body parts that do not move, such as the brain.

Several set of images are obtained during MRI imaging of a brain and the kind of imaging is dependent on clinical requirements to highlight structures or pathological processes. Brain Tumor related MRI images typically include the following:

- T1 Image (T1-weighted image)
  - T1 images are obtained by detecting signals of proton re-alignment with the magnetic field, which highlight fat in tissues of the body.
- T1Gd (or T1ce)
  - Abnormal tissue such as cancerous or inflamed tissue is often more vascular than surrounding tissue. An MRI scan can gain a clearer visual of an abnormal tissue by introducing a contrast agent. The most common contrast agent is gadolinium, a paramagnetic substance which produces very high T1 signal. A T1Gd image is a T1 image obtained after administering the contrast agent (usually intravenously). Such images enhances the abnormal tissue, where it appears brighter, than it would in pre-contrast images.
- T2 Image (T2-weighted image)
  - T2 images are obtained by detecting signals of proton spin dephasing, which highlight water and fat in tissues of the body. By comparing a T2 image with its corresponding T1 image, one can differentiate fat and water in tissues of the body.
- FLAIR (Fluid Attenuated Inversion Recovery)
  - FLAIR is a specialized MRI image, where a signal from free fluid such as cerebral spinal fluid is suppressed (compared to a T2 image). FLAIR is commonly used in brain imaging to answer specific questions related to pathological processes such as infection, tumor or areas of demyelination.

My research leverages the above MRI image sequences to develop deep learning models to identify brain tumor sub regions and mark their boundaries, as well as determine the methylation status of MGMT promoter.

# Data – Multiparametric MRI Scans [1,2,3,27]

Data for my research is acquired from several sources including the BRATS21 dataset, provided by Radiology Society of North America (RSNA), American Society of Neuroradiology (ASNR) and Medical Image Computing and Computer Assisted Interventions (MICCAI). This dataset includes retrospective collection of pre-operative multi-parametric MRI scans (mpMRI) of brain, from multiple institutions with different equipment and imaging protocols and with vastly heterogenous image quality[1]. Additionally, I sourced the images form teaching fles of local neuroradiologists and TCIA, which I used during the field testing phase.

The collected multi-parametric MRI scans consist of the following 3D image types and sequences: T1 Image (Native); T1Gd/T1Ce Image (Post Contrast); T2 Weighted Image; FLAIR

I gathered over 1500 cases for my research, from several institutions, which differ in quality and format. The data was managed in two different file formats. NIfTI (.nii) file format for segmentation and DICOM file format for radiogenomic classification of MGMT methylation status. The BRATS21 dataset also provided ground truth annotations of brain tumor sub regions, to facilitate training and cross validation. The annotated sub regions comprise the following labels: Necrotic Tumor Core (Class 1); Peritumoral edematous/invaded tissue (Class 2); Enhancing Tumor (Class 4); Image Background with no tumor (Class 0).

The annotations represent different overlapping regions of the brain tumor including enhancing tumor (class 4), tumor core (classes 2 and 4) and the whole tumor (classes 1, 2 and 4). The following diagram illustrates the different image modalities and the expert annotations, in the NIfTI file format for one case.

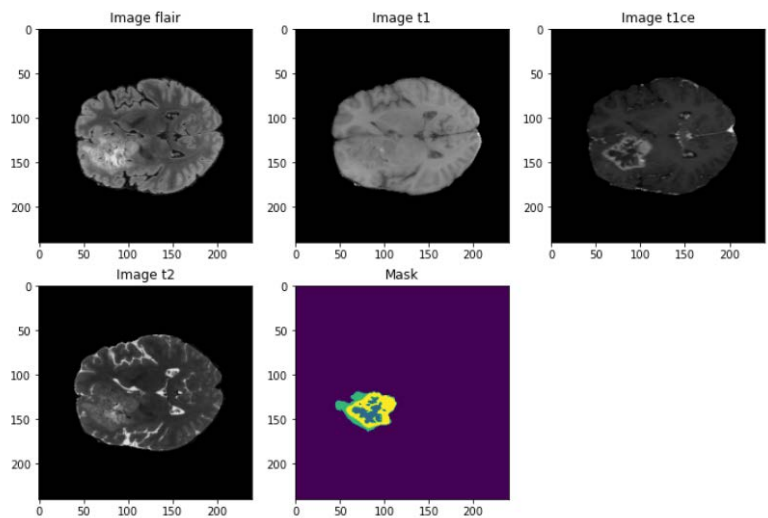

DICOM images were acquired with different slice thicknesses in Coronal, Sagittal and Axial planes. The data across the planes was not consistent for all cases. Additionally, BRATS21 dataset included a .csv file indicating the methylation status of MGMT promoter based on the laboratory assessment of surgically obtained brain tumor specimen. The methylation status is presented as a binary value of 0 (unmethylated) and 1(methylated). This data is used along with the DICOM image files to train the radiogenomic model to predict the methylation status of the MGMT promoter. The following diagram illustrates the images for different MRI modalities.

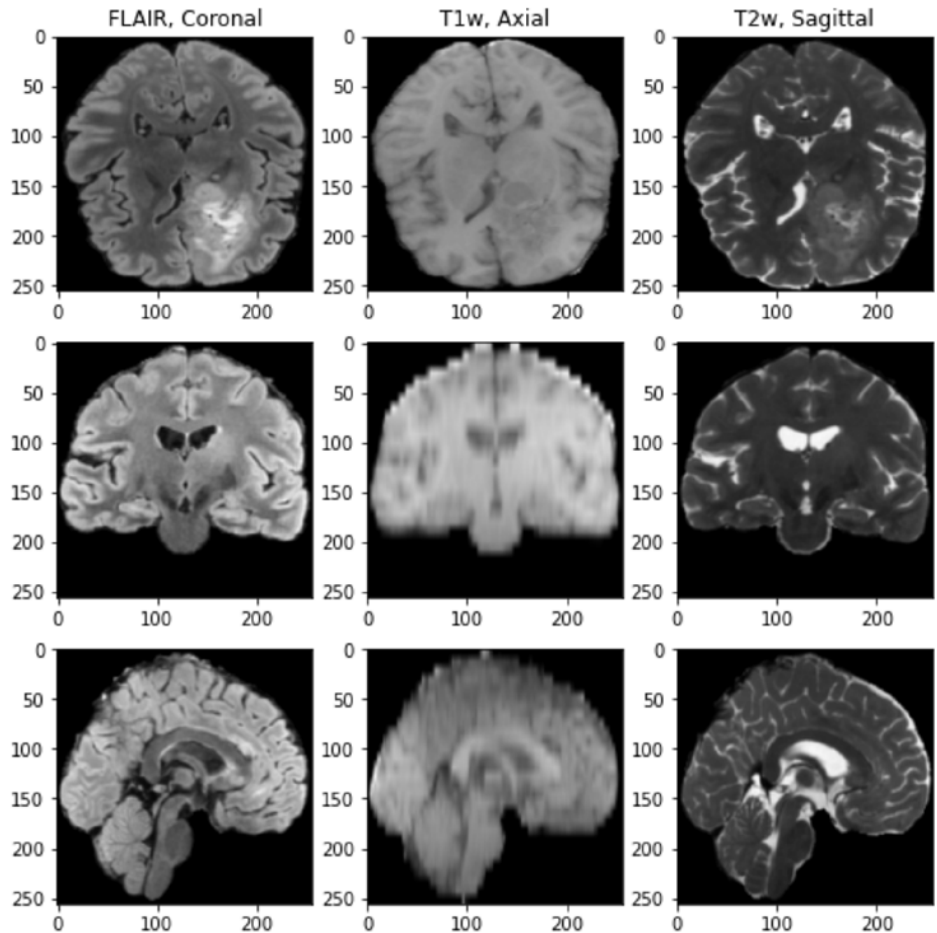

I used the BRATS21 training dataset for model training purposes, as it contained ground truth annotations of brain tumor sub regions confirmed by expert neuroradiologists, as well as methylation status of MGMT promoter conformed by laboratory assessment. The validation dataset from the BRATS21 dataset and the locally acquired images from the teaching files of board-certified neuro-radiologist did not contain ground truth annotations and hence these images were used for model testing purposes. Model predictions using the validation dataset and the locally acquired images were subjected to manual grading by local board-certified neuro-radiologist(s). I leveraged the feedback on model predictions, to further improve preprocessing and model training activities and achieved performance exceeding current standards, on both the segmentation model and the radiogenomic model. In addition, I have laid the groundwork to further my research to study additional biomarkers including IDH mutations and EGFRvIII mutations.

# Technology Framework Overview

The goal of my research is to develop an Integrated Diagnostic tool to aid physicians and brain tumor patients which can facilitate effective diagnosis, prognosis and treatment planning. For this reason, I extended my research beyond BRATS21 dataset and created a robust technology framework that provides the following:

1. A comprehensive data management component to handle data from multiple sources and formats
2. Capabilities that leverage state of the art programming frameworks and libraries to build multiple preprocessing and augmentation routines
3. Advanced computing power to experiment with several deep learning by adjusting critical parameters
4. A simple user interface to upload new images and provide prediction(s) using different models (this is particularly useful when acquiring new test images from different sources)

I conducted model training and testing on various frameworks in the Azure and Google platforms, as detailed below:

Azure Virtual Machines for Model Training, Testing and Validation

- Machine 1
  - NC8 Series
  - 1 GPU; 8 vCPUs; 56GB Memory; GPU memory of 12GB
  - Processor – NVIDIA Tesla K80
- Machine 2
  - NC12 Series
  - 2 GPUs (Tesla K80); 12 vCPUs; 112 GB Memory; GPU Memory 24 GB
  - Disk: 680 GB SSD
- Machine 3
  - NV24
  - 4 GPUs (Tesla M60); 24 vCPUs; 224 GB Memory; GPU Memory 32 GB
  - Disk: 1.44 TB SSD (local); 1 TB Azure Blob Storage

Google Colab Pro+

- 8 CPU Cores; V100 GPU and A100 GPU – 52GB RAM
- Storage – 2TB (Drive) ; 166GB Local Storage

The best performing models were trained in multiple folds, on Azure Machine 3, which is the most powerful technology framework used in my research.

Deployment (User Interface)

The AI solution is deployed using Azure Streamlit framework with a simple web application, where a physician can submit Brain MRI images in NIFTI and DICOM formats, for brain tumor segmentation and MGMT methylation prediction. Additionally, the user interface provides an option to upload ground truth annotations for comparison purposes.

The following diagrams illustrate several components and associated details of the technology and solution framework which provided the backbone for my research.

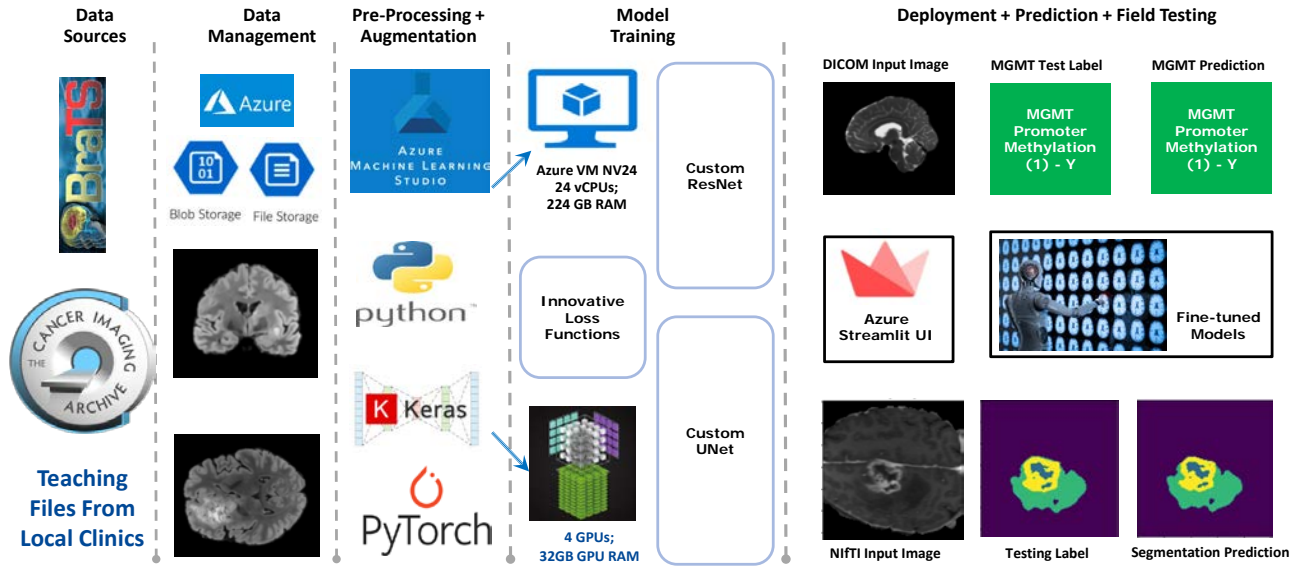

**Data Management**

- Data Gathering (routines to download and format data)
- Data Organization (folder structure for current and future data)

**Data Pre Processing and Augmentation**

- Normalization (Minmax; Zscore)
- Stack/Combine images (multi channel input)
- Crop
- Flip: Rotate
- Zoom-in: Zoom-out: Stretch
- Anisotropic Scaling
- Noise: Blur
- Volumentations: GAN
- Gretel Synthetics

**Model Training (Iterative) and Validation**

- Model Selection
  - Unet (Segmentation)
  - Simple CNN: ResNet: Effnet (MGMT Methylation Prediction)
- Network Design and Experimentation
  - Depth (Number of Layers)
  - Width (Input Size)
  - Number of filters: sizes and strides
  - Weights for classes
  - Learning Rate
  - Epochs
- Loss Function
- SGD, Adam and other variations
- Validation with public datasets
- Manual grading of test images by board certified neuroradiologist

**Deployment + Prediction**

- Azure Streamlit Framework
- User Interface to feed MRI images
- NIFTI and DICOM images for Segmentation and MGMT Methylation prediction
- Format Validation and Conversion
- Output - Segmentation of Brain tumor sub regions and Prediction of MGMT Methylation Status

**Field Testing**

- Additional Testing Data (brain MRI images) from teaching files of local neuroradiologists
- Manual Grading by board certified Neuroradiologists
- Accommodation to expand research into other use cases:
  - IDH1 mutations
  - EGFRvIII mutations

# Solution Framework and Components

## Solution Goals

My research's mission is to develop a readily deployable solution to aid physicians in effectively segmenting brain tumor sub regions as well as provide the methylation status of the MGMT promoter, a biomarker that is a favorable prognostic factor and a predictor of chemotherapy response. My research, built *segmentation model(s)* aimed at providing accurate and consistent segmentation of the brain tumor sub regions and *radiogenomic model(s)* aimed at predicting the methylation status of the MGMT promoter. Both of these solutions take multiparametric MRI scans as input. The segmentation model(s) provide a 3D image with segmentation of tumor subregions as the output, whereas the radiogenomic model(s) provide a binary classification of MGMT promoter methylation status. In addition, a stretch goal of my research is to provide a framework to extend the use cases to study additional bio markers including IDH mutations as well as EGFRvIII mutations, with MRI scans as input. My AI solution is easily extensible to accommodate the study of such additional biomarkers, which are of critical importance in guiding Glioblastoma patients with effective treatment planning and improving survival rate.

## Data Management

My research is driven by data acquired from several sources. While my solution leveraged the BRATS 2021 dataset for model training for both the segmentation model as well as the radiogenomic model, additional datasets from teaching files of local neuroradiologists were leveraged during the validation and testing phases.

To further data acquisition, in future, I created a data management framework in MS Azure platform, using Azure Blob storage, which facilitates data organization from multiple institutions and converts them to a consistent format, which can then be used in training, validation and testing phases.

## Data Pre-processing

Data pre-processing and data augmentation are necessary techniques, often leveraged in machine learning exercises, aimed at improving model performance. Several previous studies in this space have implemented both conventional and custom data pre-processing and augmentation techniques including affine, elastic and pixel level transformations[11].

The following outlines some of the pre-processing and data augmentation methods which were used in several experiments.

### Normalization

Voxel density in MRI images vary significantly from scanner to scanner as well as from case to case within a scanner, based on hardware settings. As the data was acquired form several institutions, it is necessary that images are normalized to a common scale.

I experimented with Minmax scaler technique, where the intensity is scaled to a fixed range of 0 to 1. The technique includes calculation of new scale value by subtracting the mean value from the current value and then dividing it by the current range, as outlined below:

$$x_{new} = \frac{x - x_{mean}}{x_{max} - x_{min}}$$

I also experimented with Z-score normalization where the new scale value is calculated by subtracting the mean value from the current value and dividing that by the standard deviation.

$$x_{new} = \frac{x - x_{mean}}{\sigma}$$

I found that the ultimate model performance did not differ significantly, with either of these techniques.

**Multichannel Image Input**

The multiparametric MRI scans included four different sequences, for each case, including T1, T1Ce (T1Gd), T2 and FLAIR. Both the segmentation model and the radiogenomic model would need to consider all of these sequences for tumor sub region segmentation and MGMT methylation prediction, respectively. For segmentation model(s), I used a fairly common technique to provide multiple images to the model which is to stack images (np.stack()) and create a multichannel volume for further processing and loading.

While I used the multichannel input mechanism for training the segmentation model(s), I implemented different approaches for the radiogenomic model(s), as the provided data was not consistent across all of the planes. Implemented approaches include multi-channel input as well as one channel with one image type with overall average. Ultimately, I used the input of one image type (one channel) based on the feedback from the local neuroradiologist(s), as they saw value in predicting the methylation status with each image type, independently. This technique also proved to be valuable when certain images were not available in specific studies, where the model is still capable of predicting the methylation status with greater degree of accuracy.

**Crop**

BRATS dataset provided NIfTI mages with dimensions of 240X240X155 voxels and DICOM images with dimensions of 256X256Xvariable thickness. The data acquired from teaching files of local neuroradiologists were in several different formats including 512X512Xvariable thickness. All of the acquired images were skull stripped and had anonymized by removing all of the patient's data. To gain performance in model training and also to increase overall validation efficacy, I implemented several cropping techniques including static cropping, random cropping and intelligent-cropping techniques and used the cropped section of the image for model training. I tried different cropped dimensions, both in a static and random way, and arrived at a more optimum method for model training.

For segmentation purposes, I chose a custom cropping method with a size of 128X128X128 with checks related to foreground voxels. In addition to increasing efficiency in model training, the custom cropping technique with optimum foreground voxels provided better validation performance. I also tried cropping the image to 192X192X128 image size to increase the width, but the model training times were significantly longer, with no significant improvement in performance. so I ultimately dropped this size to facilitate faster experimentation for segmentation.

For the radiogenomic model, I retained the original size of 256X256 in height and width, but chose a section of images (slices) that likely contained the tumor region, by applying different percentages determined by experiments. I then subjected the section of the slices for model training and prediction.

## Data Augmentation

With respect to data augmentation, I researched several studies and found that they provided notable performance improvements with the public datasets. For e.g., Nalepa et al., provided several studies and the impact of data augmentation on the performance of deep learning models in the brain tumor image segmentation space. Micale Futrega et al [23]., implemented several conventional techniques, to gain optimum performance of their segmentation model. Gab Allah et al., leveraged PGGAN-based augmentation to achieve higher performance of brain tumor classification model[28]. My research goes beyond the BRATS21 dataset, where my aim is to leverage my model to segment brain tumor sub regions and predict MGMT methylation status for general MRI scans, at local healthcare institutions. While I was interested in gaining higher performance with the BRATS21 dataset, I also wanted models that can perform better, under manual grading by board certified neuroradiologists, with real world MRI scans from local institutions. For this reason, I experimented with several conventional data augmentations such as flip, rotate and stretch, as well as custom augmentations including Generating Adversarial Networks (GAN), to induce new data.

As expected, models trained with induced data provided initial performance improvements with public training datasets, but the cross-validation performance with validation and testing datasets deteriorated. Moreover, when the model was fed data from teaching files of local neuroradiologists, the performance deteriorated even further. Contrarily, the models trained with limited conventional augmentations, such as rotate, flip and stretch or no augmentations at all, performed better with validation and test datasets, as well as with the images from the teaching files of local neuroradiologists.

As my research's goal is to apply my solution to any brain MRI image (not just the BRATS21 dataset), I chose to use the models with robust network design and relied on optimum training to gain the best performance rather than using models that were trained on augmented datasets which performed better on public training datasets and not very well on validation and testing datasets and even worse on the studies from the teaching files of local neuroradiologists. While I retained the pre-processing steps across several models, I dropped the custom data augmentation techniques which induce data on the models I chose for final testing.

## Model Design and Training

A well-designed network with optimized parameters is crucial for achieving a deep learning solution that delivers high-performance deep. I have conducted several experiments, over 75 experiments using several CNNs for both segmentation and radiogenomic models. I documented the performance metrics of some of the high-performance models in the "Results and Prediction" section. More importantly I developed a framework to quickly test existing and new data using the high-performance models, with a goal to provide an Integrated Diagnostics solution to aid physicians with effective diagnosis, prognosis and treatment planning for brain tumor patients. The following describes several key aspects related to model selection, network design and training techniques for both radiogenomic and segmentation models.

**Radiogenomic model**
The goal of the radiogenomic model is to understand the genetic and molecular characteristics in the set of input images using CNN feature extractors, and predict the methylation status of MGMT (Methylguanine Methyltransferase) in it is promoter region. MGMT is a DNA repair enzyme that plays an important role in chemoresistance to alkylating agents. Methylation of MGMT promoter is a key predictor of whether alkylating agents can effectively control glioma cells. Current methods of predicting MGMT methylation required surgical biopsies, at times multiple. These surgical procedures are time consuming and not all patients are candidates for surgical biopsies. There are also issues related to accuracy of MGMT methylation determination using surgical biopsies as it largely depends on the location of the biopsy. Hence a non-invasive means to determine the MGMT methylation status, more accurately, is immensely useful in planning treatment for brain tumor patients. While segmentation model provides important information to monitor disease progression and aid in other clinical applications, it is the radiogenomic model that plays a crucial role in determining the patient's response to chemotherapy and accordingly guide the patients to chemotherapy if they have methylated MGMT or more importantly guide them to other treatments if they have unmethylated MGMT.

I conducted several experiments using simple CNNs from Keras, MONAU and other PyTorch libraries. I used models with pre-trained weights as well as vanilla models. I tried several architectures including ResNet10, ResNEt 50, ResNEt101 and EffNEt architectures. I experimented with different optimizers including SGD and ADAM optimizer for this models. I also used different learning rates 0.0001, 0.00005 and 0.00001 (which varied based on the network and the number of epochs), and chose faster learning rates for earlier epochs and slower learning rates for later epochs. Several of these models performed will with the public dataset provided by BRATS21 and the best performing model with this training dataset achieved a validation AUROC of 94% with specific image types. However, as previously noted, the best performing model (a 2D model) with the training dataset did not yield acceptable results when applied on the images from the teaching files of a local neuroradiologist. As I expanded my experiments, I found that 3D models with no significant augmentations performed the best with different datasets.

Consequently, I chose a robust 3D model, based on ResNet architecture. I implemented Binary Cross Entropy los function and ADAM optimizer with variable learning rate. I implemented a slice selection technique that choses the slices with bulk of the tumor and ignore the slices that do not have the tumor. After experimentation, I found that the slices that are in the 25%-75% range, provided the best performance and accordingly used this selection for model training and inference. With training and continuous improvement, I was able to achieve an overall average performance of 83% AUROC, which exceeded the winning model by Faris Baba, in 2021-Kaggle competition, by 21 points. The model performance is even better, 85%, when MGMT methylation status is predicted with T1WCE image. This model also achieved consistently superior results, on several brain MRI studies from both public datasets and private studies. **With performance exceeding the Kaggle winning model by 21 points, my radiogenomic model to predict MGMT methylation status is easily the biggest differentiator of my research.** The following diagrams show the network design and key design parameters of my best performing model.

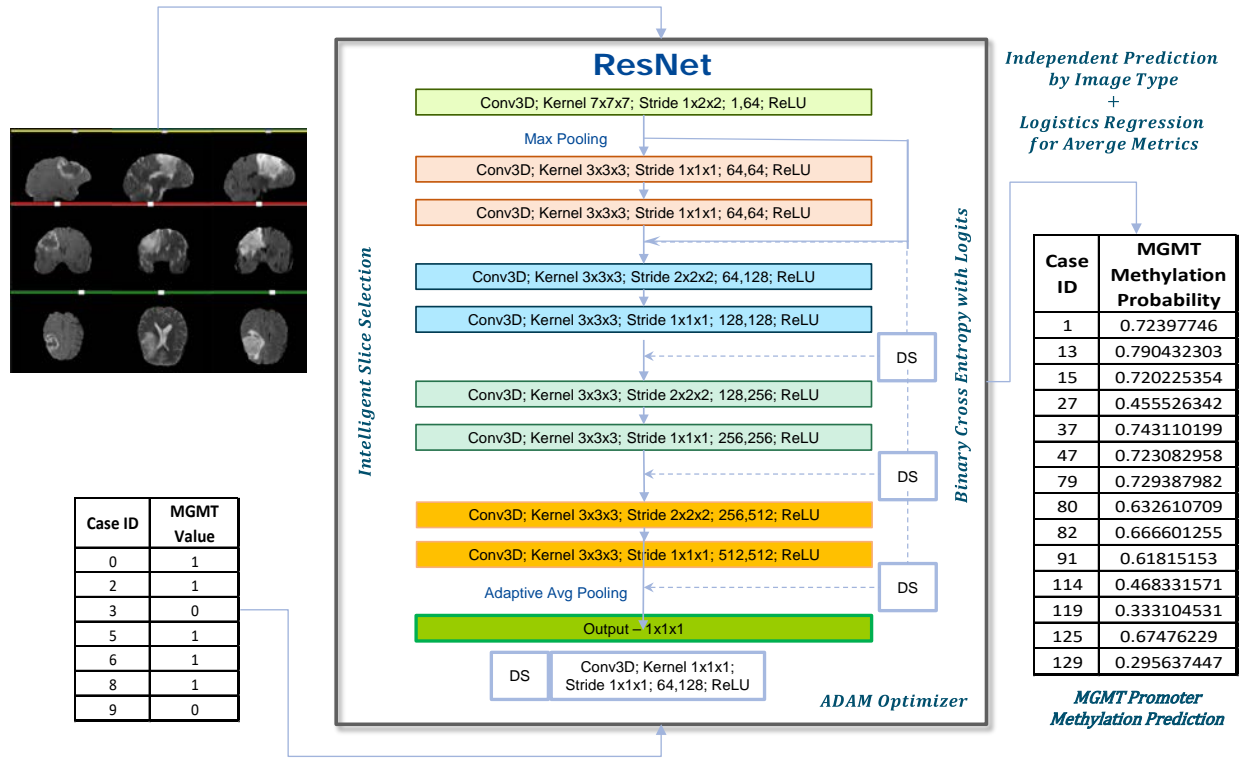

I conducted 4-fold model training and evaluated AUROC for each image type independently, where the data was split with 80-20 ratio, for training and validation. The model performance was higher with training and validation data, for e.g. it achieved a 95% validation AUROC with T1WCE. I then performed a logistics regression with 16 different weights (4 weights, one for each image type, for each of the 4 folds). Subsequently, I subjected my model to predict MGMT methylation status on unseen testing data. With unseen data, the model provided 83% overall AUROC and 85% with T1WCE.

## Segmentation Model

Deep learning-based image processing typically deals with two different kinds of problems: Classification and Segmentation. In classification problems the model is trying to predict what is in the image and classifies it in alignment with the prediction criteria. Segmentation is different from classification in the sense that the model is expected not only to predict what the target image (which is usually a portion of the input image) is but also provide where the target image is. Often the segmentation problems include identifying several target images in the input image as well as drawing boundaries around each target object to identify the location of the target image. In some ways, segmentation problem is a classification problem at a pixel/voxel level, which classifies the target objects and draws boundaries around them to identify where they are. The brain tumor segmentation problem is a typical segmentation problem, where the objective is to identify several sub regions of the tumor and draw boundaries around them to segment one from another. While there are several convolutional neural networks which aid in solving segmentation problems, research has shown that a well-designed U-Net model provides the best results, particularly in segmenting medical images.

For my research I chose a vanilla U-net model with custom input parameters, kernel dimensions and depth to segment brain tumor sub regions. During experimentation, I have attempted several different depths including 5, 7 and 9 layers and ultimately chose a simpler model with 7 layers, which performed better and the training time was faster. While the network with 9 layers performed well, there was a slight deterioration of validation metrics with the 9-layer network and the training time was also significantly longer. I experimented with different input parameters including 3 channels (T1CE; T2 and FLAIR) as well as 4 channels (T1; T2; T1CE; FLAIR), to determine if T1 image contributed any significant value towards tumor segmentation. I found that the performance was nearly similar with and without the T1 image. While I did retain the T1 image in the input for my model, I wanted to measure the contribution of T1 image and hence the reason for experimentation. I customized the number of convolutions, number of filters and at times the strides of the filters at each level and arrived at the most optimum network design for the final model.

I experimented with several loss functions, including Combo Loss (Dice Loss + Weighted Focal Loss) and Unified Focal Loss. The model performance during training and validation was different with each loss function, but they were within acceptable range. However, the model trained with Combo Loss provided best field testing results and accordingly, I retained this loss function for deployment. I experimented with several Optimizers (SGD; Adam, and variations of Adam) and ultimately chose Adam which delivered the highest performance. I also experimented with different learning rates with 0.0001 as the default learning rate for lower numbered epochs and then used 0.00005 and 0.000001 for later epochs. The learning rate did not seem to make much of a difference and consequently I retained 0.0001 as the default rate. I have achieved acceptable performance with several models, ranging from 85% to 95% in Dice Score, with the training datasets. The highest performance of 95% was achieved, when training with significantly augmented data. This performance is comparable to the winning model by Futrega et al.[22], in the BRATS 2021 competition. However, based on the feedback from neurosurgeons and neuroradiologist(s), on the performance of several models on locally acquired images, which were significantly different in quality and format, with additional analysis of the nature of false positives and the sensitivity with which the tumor core and enhancing tumor were segmented, I determined that the model which was trained without significant augmentations, performed better overall, when applied to unseen MRI studies with varying degrees of quality. Accordingly, I reduced the custom augmentations to simulate real world MRI scans and I was able to achieve models which had slightly lesser Dice Score 86% with the training dataset, but performed exceedingly well under manual grading with locally acquired images as well as with the validation dataset from BRATS21. The results of the best performing model, so far, are outlined in the subsequent sections. The following diagram outlines the network design and key parameters of the Unet model with 7 layers.

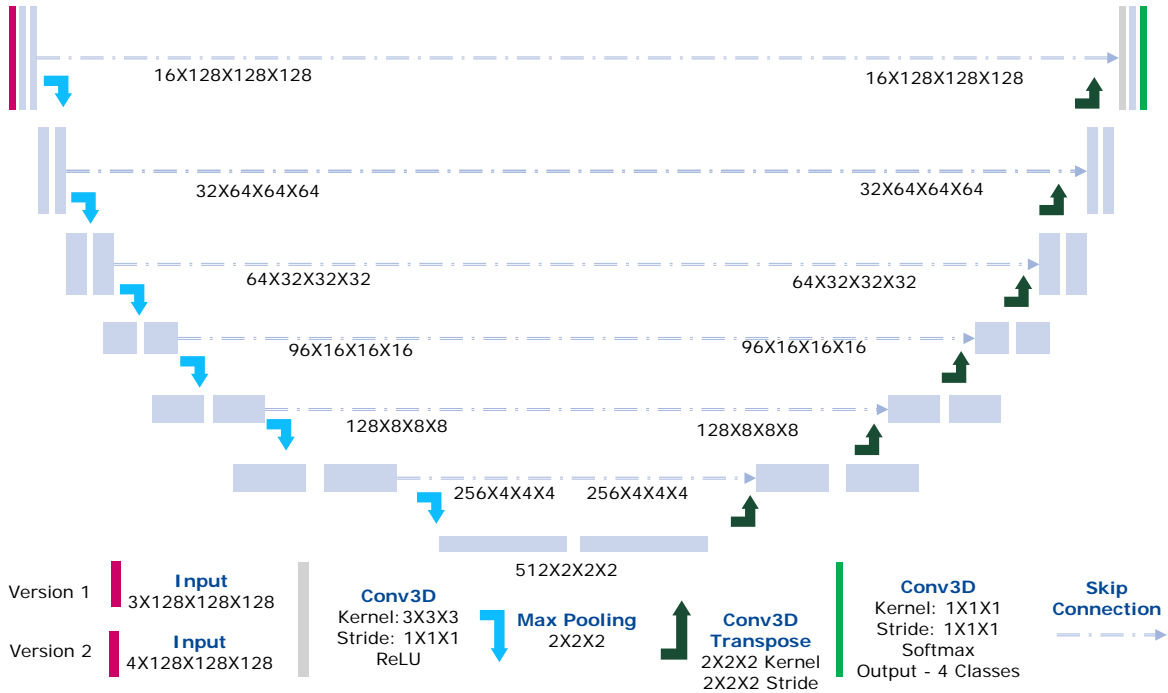

The following diagram outlines key design parameters and some predictions of my segmentation model.

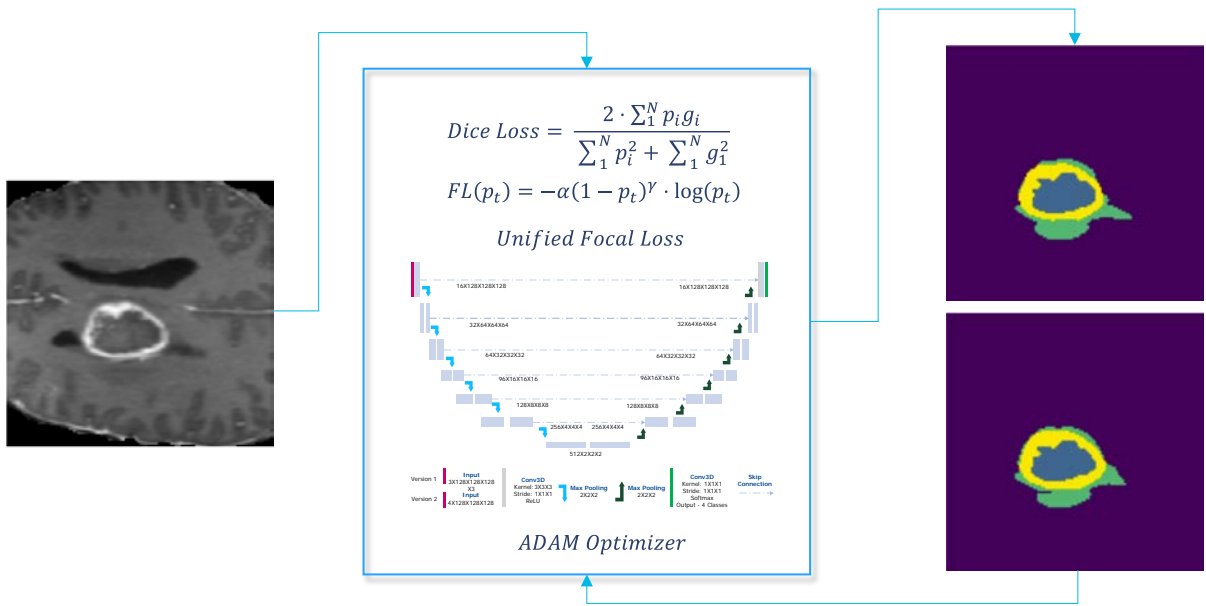

I chose a 3D model for final deployment as it performed best in field testing. I also experimented with 2D models, as prior research (and my own experiments) has shown that 2D models provided better performance with training and validation on public datasets. However, I found that the random distribution of images in the 2D model may split the data in such a way that the training and validation sections may have the similar slices of the same image and accordingly I realized that the performance metrics of 2D models are artificially high. This is supported by less than acceptable performance of the 2D models in field testing. Hence, I chose to retain 3D models for field testing and final deployment.

## Model Performance and Predictions

### Performance metrics of a better performing Radiogenomic Model(s)

The metrics shown below are from a 2D model that performed significantly better under the training dataset from BRATS21 but did not perfrom as well when applied to the images from the teaching files of local neuroradiologist. This model provided an near perfect training AUROC and over 94% validation AUCROC. However, it did not perform well on testing datasets.

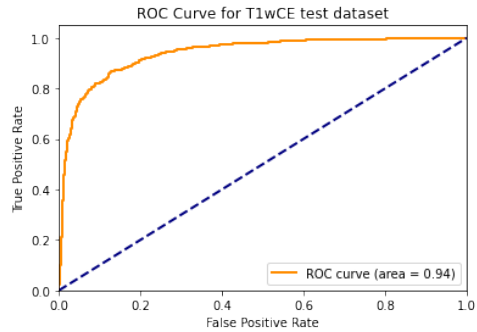

As noted previously, another representative model, shown below, performed within an acceptable range with the BRATS21 training dataset, with validation AUROC score of ~83%, as shown below. This score well exceeds the winning model from Kaggle, and it also performed significantly better under manual grading with the images form the teaching files of local neuroradiologists. Accordingly, I retained this model for final deployment.

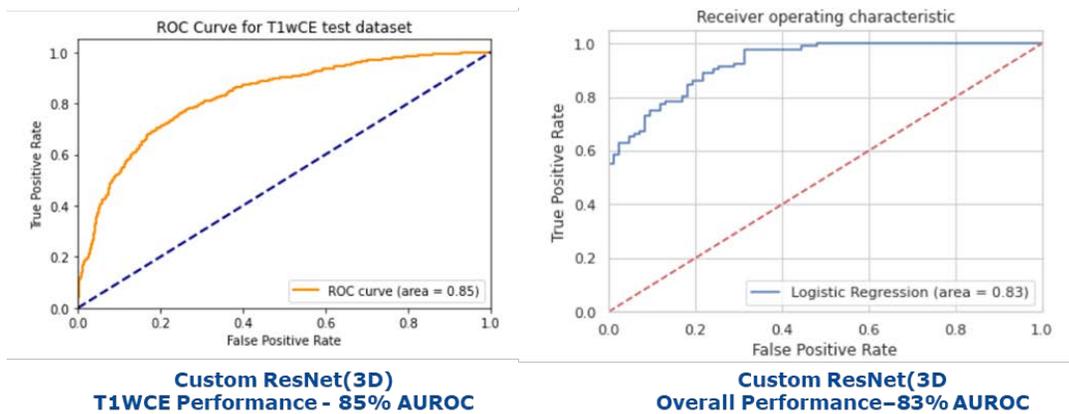

**Custom ResNet(3D)**
**T1WCE Performance - 85% AUROC**

**Custom ResNet(3D**
**Overall Performance~83% AUROC**

**Performance metrics of a better performing Segmentation Model (U Net-7)**

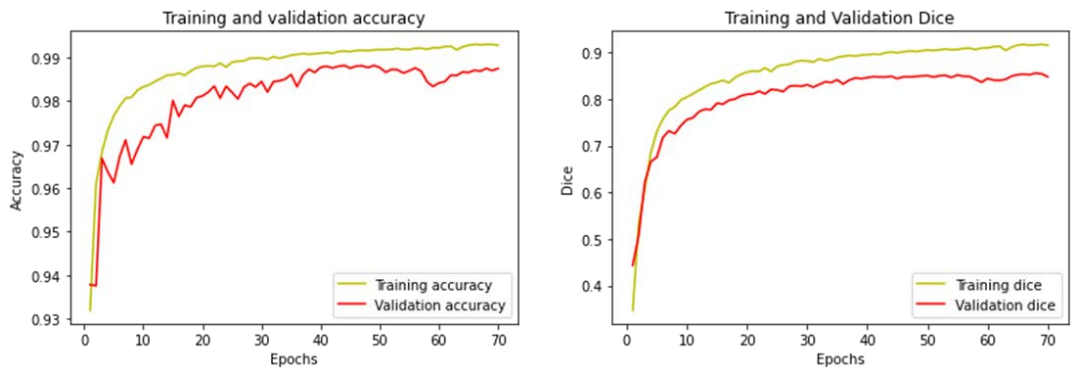

The above model provided a near perfect training and validation accuracy of ~99%, with a traning dice score of ~92% and a validation dice score of ~86%.

The following diagram shows model prediction of segmented subregions of brain tumor for an example test case. The model performance was comparable under manual grading with some scans in the validation and testing datasets.

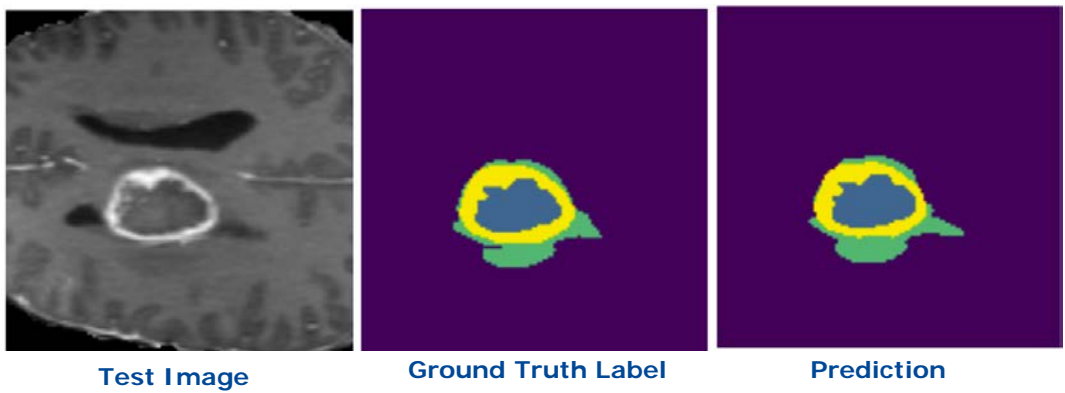

Test Image          Ground Truth Label          Prediction

One of my experiments with data augmentations, provided superior training and validation performance of Dice score of ~95%, which is better than the winning model in BRATS21 competition. However, the performance of the above model had several false postives during the field testing phase and accordingly I chose the best performing model with accuracy of ~99% and dice score of ~86%.

The following diagrams show model predictions during the field testing with images from the teaching files of local neuroradiologists. These images were significanlty different from the BRATS21 dataset, both in quality and format. The model predictions provided near perfect segmentation of brain tumor subregions, confirmed with manual grading by local neuroradiologists.

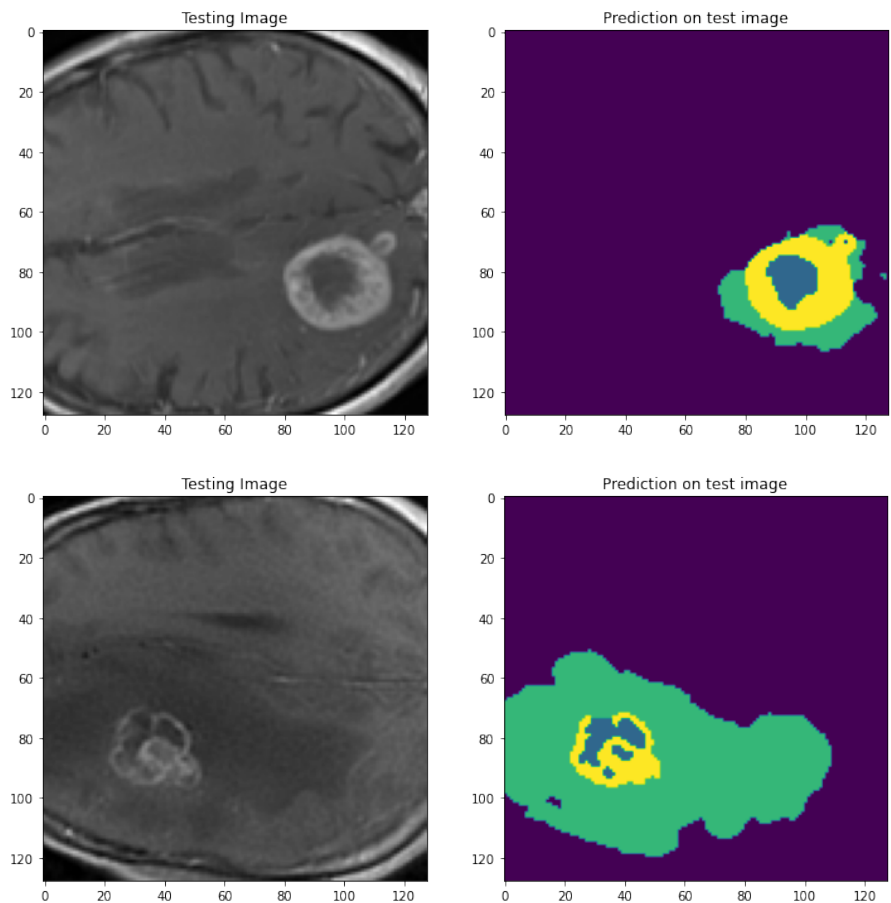

## Conclusions

The primary goal of my research is to develop an integrated diagnostic tool powered by deep learning models, to aid physicians and brain tumor patients which can facilitate effective diagnosis, prognosis and treatment planning and ultimately improve patient survival time. While more research is needed before putting such solutions in practice, initial results from my experiments prove very promising and encouraging.

I experimented with several deep learning models for both brain tumor segmentation and methylation classification of MGMT promoter, and found that simpler models with relatively smaller number of layers performed better than models with higher number of layers (more depth). In other words complexity and depth of the models did not add value and they performed less effectively than the simpler models.

I experimented with several data augmentation techniques including conventional affine transformations that do not alter the image, as well as with techniques that induce or impute data using Generating Adversarial Networks. I found that by inducing data, there is indeed initial improvement of performance improvements with the training dataset. However, models trained with such augmented data did not perform necessarily well with testing datasets or with files from local institutions, which are beyond BRATS21 datasets. Models trained with data that is not induced provided better overall performance, especially when applied to the new datasets. We found this pattern even with the winning models in Kaggle and BRATS21 competitions. In summary, for deep learning models dealing with medical images, my research found that significant data augmentations where new data is induced may not yield better overall performance.

I experimented with several pre-trained models, using transfer learning techniques and found that they did not yield optimum performance. While I retained some of them for continued testing in my overall solution framework, it is evident from my research that vanilla networks performed much better than pre-trained models.

I experimented with both 2D and 3D models for both MGMT methylation and segmentation. While 2D models performed better in training and validation, they suffered during testing. With additional research, I determined that with 2D models, slices from the same study could be present in both training and validation and consequently the model may achieve boosted training and validation metrics. However, during field testing, such an advantage does not exist and consequently 2D models suffered significantly during field testing. 3D models on the other hand required more tuning during training to achieve comparable performance. However they performed significantly better in field testing and consequently, I only retained 3D models for final deployment.

My research also provided evidence and encouragement that my solution framework can be extended to study additional bio markers, such as EGFRvIII and IDH1 mutations, which can increase the precision with which one can predict patient's response to chemotherapy treatment and aid patients and physicians with effective treatment planning and improve patient's survival time.

## Acknowledgements

I would like to express my sincere gratitude to Dr. Venkatramana (Venkat) Vattipally, a board certified Neuroradiologist in Saginaw, Michigan. Dr. Venkat completed his fellowship in Neuroradiology from Yale-New Haven Hospital and he is currently practicing in Saginaw, MI, with affiliations to several local hospitals and clinics. My research significantly benefited from the MRI studies, acquired from the teaching files of local neuroradiologists, and provided by Dr. VenkatramanaVattipally. In addition, Dr. Venkat continues to provide feedback on the model performance of various brain tumor segmentation and radiogenomic models developed by me, which has played an immense role in improving model performance, particularly with datasets that are beyond BRATS21 training datasets. I am forever grateful to Dr. Venkat for his counsel and help with my research.

# References:

1. The RSNA-ASNR-MICCAI BraTS 2021 Benchmark on Brain Tumor Segmentation and Radiogenomic Classification - arXiv:2107.02314v2 [cs.CV] 12 Sep 2021

2. The Multimodal Brain Tumor Image Segmentation Benchmark - EEE Trans Med Imaging. 2015 October ; 34(10): 1993–2024. doi:10.1109/TMI.2014.2377694

3. Advancing The Cancer Genome Atlas glioma MRI collections with expert segmentation labels and radiomic features - Spyridon Bakas1,2, Hamed Akbari1,2, Aristeidis Sotiras1,2, Michel Bilello1,2, Martin Rozycki1,2, Justin S. Kirby3, John B. Freymann3, Keyvan Farahani4 & Christos Davatzikos1,2 - www.nature.com/scientificdata - SCIENTIFIC DATA | 4:170117 | DOI: 10.1038/sdata.2017.117

4. Current Clinical Brain Tumor Imaging - Javier E. Villanueva-Meyer, MD

5. Marc C. Mabray, MD Soonmee Cha, MD - www.neurosurgery-online.com - Neurosurgery 81:397–415, 2017 - DOI:10.1093/neuros/nyx103

6. Brain tumor segmentation based on deep learning and an attention mechanism using MRI multi-modalities brain images - www.nature.com/scientificreports - https://doi.org/10.1038/s41598-021-90428-8 - Scientific Reports | (2021) 11:10930

7. Fully Automated Brain Tumor Segmentation and Survival Prediction of Gliomas using Deep Learning and MRI - https://doi.org/10.1101/760157 - Chandan Ganesh Bangalore Yogananda, M.S.1, Sahil S. Nalawade, M.S.1, Gowtham K. Murugesan, M.S.1, Ben Wagner, B.M.1, Marco C. Pinho, M.D.1, Baowei Fei, Ph.D. 2, Ananth J. Madhuranthakam, Ph.D. 1, Joseph A. Maldjian, M.D.1, Advanced Neuroscience Imaging Research Lab, Department of Radiology, University of Texas Southwestern Medical Center, Dallas, Texas, USA, Department of Bioengineering, University of Texas at Dallas, Texas, USA.

8. The Multimodal Brain Tumor Image Segmentation Benchmark (BRATS) - Bjoern Menze, Andras Jakab, Stefan Bauer, Jayashree Kalpathy-Cramer, Keyvan Farahani, Justin Kirby, Yuliya Burren, Nicole Porz, Johannes Slotboom, Roland Wiest, et al. - Bjoern Menze, Andras Jakab, Stefan Bauer, Jayashree Kalpathy-Cramer, Keyvan Farahani, et al.. The Multimodal Brain Tumor Image Segmentation Benchmark (BRATS). IEEE Transactions on Medical Imaging, Institute of Electrical and Electronics Engineers, 2014, 34 (10), pp.1993-2024. 10.1109/TMI.2014.2377694. hal-00935640v2 - HAL Id: hal-00935640 https://hal.inria.fr/hal-00935640v2

9. Multiscale CNNs for Brain Tumor Segmentation and Diagnosis - Liya Zhao and Kebin Jia - Multimedia Information Processing Group, College of Electronic Information & Control Engineering, Beijing University of Technology, Beijing, China - Hindawi Publishing Corporation - Computational and Mathematical Methods in Medicine Volume 2016, Article ID 8356294, 7 pages - http://dx.doi.org/10.1155/2016/8356294

10. Data Augmentation for Brain-Tumor Segmentation: A Review - Jakub Nalepa 1,2*, Michal Marcinkiewicz 3 and Michal Kawulok 2 - Future Processing, Gliwice, Poland, 2 Silesian University of Technology, Gliwice, Poland, 3 Netguru, Poznan, Poland - published: 11 December 2019 doi: 10.3389/fncom.2019.00083 - Frontiers in Computational Neuroscience | www.frontiersin.org - December 2019 | Volume 13 | Article 83

11. Segmenting Brain Tumor Using Cascaded V-Nets in Multimodal MR Images - Rui Hua 1,2, Quan Huo 2, Yaozong Gao 2, He Sui 3, Bing Zhang 4, Yu Sun 1, Zhanhao Mo 3* and Feng Shi 2* - 1 School of Biological Science and Medical Engineering, Southeast

University, Nanjing, China, 2 Shanghai United Imaging Intelligence, Co., Ltd., Shanghai, China, 3 China-Japan Union Hospital of Jilin University, Changchun, China, 4 Department of Radiology, Affiliated Drum Tower Hospital of Nanjing University Medical School, Nanjing, China - ORIGINAL RESEARCH published: 14 February 2020 doi: 10.3389/fncom.2020.00009 - Frontiers in Computational Neuroscience | www.frontiersin.org - February 2020 | Volume 14 | Article 9

12. The Federated Tumor Segmentation (FeTS) Challenge - Pati et al. - arXiv:2105.05874v2 [eess.IV] 14 May 2021 - https://www.fets.ai/ - https://www.cbica.upenn.edu/captk -

13. Clinically Relevant Imaging Features for MGMT Promoter Methylation in Multiple Glioblastoma Studies: A Systematic Review and Meta-Analysis - C.H. Suh, X H.S. Kim, X S.C. Jung, X C.G. Choi, and X S.J. Kim - Published July 12, 2018 as 10.3174/ajnr.A5711- AJNR Am J Neuroradiol ●:●●2018 www.ajnr.org

14. MRI-Based Deep-Learning Method for Determining Glioma MGMT Promoter Methylation Status - C.G.B. Yogananda, B.R. Shah, S.S. Nalawade, G.K. Murugesan, F.F. Yu, M.C. Pinho, B.C. Wagner, B. Mickey, T.R. Patel, B. Fei, A.J. Madhuranthakam, and J.A. Maldjian - http://dx.doi.org/10.3174/ajnr.A7029 - JNR Am J Neuroradiol 2021 www.ajnr.org

15. Radiomics and MGMT promoter methylation for prognostication of newly diagnosed glioblastoma - www.nature.com/scientificreports - Received: 4 January 2019 Accepted: 20 September 2019 Published: 08 Oct 2019 - 14435 | https://doi.org/10.1038/s41598-019-50849-y

16. Identifying the Best Machine Learning Algorithms for Brain Tumor Segmentation, Progression Assessment, and Overall Survival Prediction in the BRATS Challenge - arXiv:1811.02629v3 [cs.CV] 23 Apr 2019 - s.bakas@uphs.upenn.edu, bjoern.menze@tum.de

17. Automatic Brain Tumor Segmentation Based on Cascaded Convolutional Neural Networks With Uncertainty Estimation - Guotai Wang 1,2*, Wenqi Li 2,3, Sébastien Ourselin 2 and Tom Vercauteren 2 - 1 School of Mechanical and Electrical Engineering, University of Electronic Science and Technology of China, Chengdu, China, 2 School of Biomedical Engineering and Imaging Sciences, King's College London, London, United Kingdom, 3 NVIDIA, Cambridge, United Kingdom - Frontiers in Computational Neuroscience | www.frontiersin.org - August 2019 | Volume 13 | Article 56

18. Learning rich features with hybrid loss for brain tumor segmentation - Daobin Huang1,2,3, Minghui Wang1, Ling Zhang4, Haichun Li1, Minquan Ye1,3* and Ao Li1* - Huang et al. BMC Med Inform Decis Mak 2021, 21(Suppl 2):63 https://doi.org/10.1186/s12911-021-01431-y - From International Conference on Health Big Data and Artificial Intelligence 2020 Guangzhou, China. 29 October - 1 November 2020

19. A Survey of MRI-Based Brain Tumor Segmentation Methods - TSINGHUA SCIENCE AND TECHNOLOGY ISSNll1007-0214ll04/10llpp578-595 Volume 19, Number 6, December 2014

20. RMU-Net: A Novel Residual Mobile U-Net Model for Brain Tumor Segmentation from MR Images - Muhammad Usman Saeed 1,* , Ghulam Ali 1 , Wang Bin 2 , Sultan H. Almotiri 3, Mohammed A. AlGhamdi 3 , Arfan Ali Nagra 4, Khalid Masood 4 and Riaz ul Amin 1 - Electronics 2021, 10, 1962. https://doi.org/10.3390/electronics10161962 - https://www.mdpi.com/journal/electronics

21. Reducing the Hausdorff Distance in Medical Image Segmentation with Convolutional Neural Networks - Davood Karimi, and Septimiu E. Salcudean, Fellow, IEEE - arXiv:1904.10030v1 [eess.IV] 22 Apr 2019

22. Optimized U-Net for Brain Tumor Segmentation - Micha l Futrega, Alexandre Milesi, Micha l Marcinkiewicz, Pablo Ribalta - arXiv:2110.03352v1 [eess.IV] 7 Oct 2021

23. Clinical applications of artificial intelligence and radiomics in neuro-oncology imaging - Ahmed Abdel Khalek Abdel Razek1, Ahmed Alksas2, Mohamed Shehata2, Amr AbdelKhalek3, Khaled Abdel Baky4, Ayman El-Baz2 and Eman Helmy1 - Abdel Razek et al. Insights Imaging (2021) 12:152 - https://doi.org/10.1186/s13244-021-01102-6

24. Aggregation-and-Attention Network for brain tumor segmentation - Chih-Wei Lin1,2,3,4* , Yu Hong2,4 and Jinfu Liu1,2,4- Lin et al. BMC Med Imaging (2021) 21:109 https://doi.org/10.1186/s12880-021-00639-8

25. Gaillard, F., Weerakkody, Y. Glioblastoma, IDH-wildtype. Reference article, Radiopaedia.org. (accessed on 27 Dec 2021) https://doi.org/10.53347/rID-4910 https://doi.org/10.53347/rID-4910

26. Segmentation Labels and Radiomic Features for the Pre-operative Scans of the TCGA-GBM collection (07 2017). https://doi.org/10.7937/K9/TCIA.2017.KLXWJJ1Q , Bakas, S., Akbari, H., Sotiras, A., Bilello, M., Rozycki, M., Kirby, J., Frey-mann, J., Farahani, K., Davatzikos, C.

27. Classification of Brain MRI Tumor Images Based on Deep Learning PGGAN Augmentation - Gab Allah, A.M.; Sarhan, A.M.; Elshennawy, N.M. Classification of Brain MRI Tumor Images Based on Deep Learning PGGAN Augmentation. Diagnostics 2021, 11, 2343. https://doi.org/ 10.3390/diagnostics11122343.

28. Optimizing Prediction of MGMT Promoter Methylation from MRI Scans using Adversarial Learning - Sauman Das - arXiv:2201.04416v1 [eess.IV] 12 Jan 2022

29. Glioblastoma patient statement from Gliocure – A research oriented organization - https://www.gliocure.com/en/patients/glioblastoma/